\documentclass[prl,twocolumn,nofootinbib, preprintnumbers, superscriptaddress]{revtex4-1}

\usepackage{amsmath,amssymb,amscd,simplewick}
\usepackage{listings}
\usepackage{dsfont}
\usepackage{slashed}
\usepackage{color}
\usepackage{ulem}

\usepackage{graphicx}
\usepackage{epstopdf}
\usepackage{subfigure}
\usepackage{epsfig}

\usepackage{xcolor}
\usepackage[colorlinks=true,
            linkcolor=blue,
            urlcolor=blue,
            citecolor=green,          
            bookmarks=true,
            bookmarksnumbered=true,
            breaklinks=true,
            pdfpagemode=Fullscreen,
            pdfstartview=FitBH]{hyperref}

\hypersetup{pdfauthor = {Shao-Feng Ge},
	     pdftitle = {}, 
	     pdfsubject = {}, 
             pdfkeywords = {}, 
	     pdfcreator = {LaTeX with hyperref package},
	     pdfproducer = {dvips + ps2pdf} }

\definecolor{gesfpurple}{rgb}{0.47,0.19,0.42}

\definecolor{gesflanse}{rgb}{0.00,0.50,0.50}

\definecolor{gesfblue}{rgb}{0.08,0.42,0.76}

\definecolor{gesfred}{rgb}{1,0,0}

\definecolor{gesfwhite}{rgb}{1,1,1}

\definecolor{gesfblack}{rgb}{0,0,0}

\newcommand{\geqn}[1]{\hypersetup{linkcolor=blue}(\ref{#1})\hypersetup{linkcolor=blue}}
\newcommand{\gfig}[1]{{\hypersetup{linkcolor=violet}Fig.~\ref{#1}\hypersetup{linkcolor=blue}}}

\newcommand{\fp}{\slashed p}

\graphicspath{{figs/}}

\begin{document}

\preprint{\today \hspace{14cm} IPMU19-0047}

\title{\fontsize{11pt}{13pt}\selectfont Apparent CPT Violation in Neutrino Oscillation from Dark Non-Standard Interactions}

\vskip 0.5in
\author{Shao-Feng Ge}
\email{gesf02@gmail.com}
\affiliation{Kavli IPMU (WPI), UTIAS, University of Tokyo, Kashiwa, Chiba 277-8583, Japan}
\affiliation{Department of Physics, University of California, Berkeley, CA 94720, USA}
\affiliation{T.~D.~Lee Institute, Shanghai 200240, China}
\affiliation{School of Physics and Astronomy, Shanghai Jiao Tong University, Shanghai 200240, China}
\author{Hitoshi Murayama}
\email{hitoshi@berkeley.edu, hitoshi.murayama@ipmu.jp}
\affiliation{Department of Physics, University of California, Berkeley, CA 94720, USA}
\affiliation{Kavli IPMU (WPI), UTIAS, University of Tokyo, Kashiwa, Chiba 277-8583, Japan}
\affiliation{Ernest Orlando Lawrence Berkeley National Laboratory, Berkeley, CA 94720, USA}

\begin{abstract}
A natural realization of CPT violation in neutrino oscillation can arise
due to the coupling to a light scalar or vector dark matter (DM).
The dark non-standard interaction (NSI) is associated with the $\gamma_0$ matrix in neutrino's
effective propagator and hence corrects the neutrino Hamiltonian as {\it dark}
matter potential, in the same way as the ordinary matter effect.
The effect is, however, inversely proportional to the neutrino
energy and hence appears as a correction to the neutrino mass squared.
Due to a sign
difference in the corrections for neutrino and anti-neutrino modes,
the neutrino oscillation receives CPT violation
from the dark NSI. Seeing difference
in the neutrino and anti-neutrino mass squared differences not necessarily
leads to the conclusion of CPT symmetry breaking in the fundamental Lagrangian
but can indicate light DM and its coupling with neutrinos.
\end{abstract}

\maketitle 

{\it Introduction} --
The CPT theorem is one of a few robust predictions of the relativistic local quantum
field theories (QFT) \cite{Streater:1989vi,RalfLehnert:2016grl}. As long as a theory
satisfies three conditions: 1) Lorentz invariance, 2) hermiticity of the Hamiltonian,
and 3) locality, it is invariant under the combined CPT transformation.
The CPT violation then unavoidably indicates violation of at least
one of the three conditions. Note that these three conditions are quite
fundamental and measuring CPT violation is a direct probe of the underlying
structure of the Nature.

A phenomenological consequence of the CPT symmetry is that
a particle and its anti-particle must have exactly the same mass and lifetime.
Measuring the difference in the particle and anti-particle masses and lifetimes is then
a direct probe of the CPT symmetry.
This applies to the neutral Kaon and neutrino systems.
Although the constraint from the neutral Kaon system seems quite stringent,
$|m(K^0) - m(\overline K^0)| / m_K < 6 \times 10^{-18}$ \cite{PDG18},
a more natural parametrization is in terms of the mass squared.
First, the parameter that appears in the Lagrangian is $m^2_K$ rather than $m_K$,
Even for the neutrino system,
although the fermion mass appears as $m_\nu$ in the Lagrangian, it is the mass
squared terms in the Hamiltonian that control the oscillation pattern.
Using the mass squared parametrization, the Kaon constraint
$|m^2(K^0) - m^2(\overline K^0)| < 0.25~\mbox{eV}^2$, reads much weaker
and the neutrino system actually
gives better constraint \cite{Murayama:2003zw}.

Neutrinos are more fundamental particles than the neutral Kaons and hence are
probably better probes of the fundamental CPT symmetry \cite{Barenboim:2017ewj}.
Currently neutrino oscillation provides the most stringent bound,
$|\Delta m^2_{21} - \Delta \overline m^2_{21}| < 5.9 \times 10^{-5} \mbox{eV}^2$
and
$|\Delta m^2_{31} - \Delta \overline m^2_{31} | < 1.1 \times 10^{-3} \mbox{eV}^2$
\cite{CPT-osc}.
The future DUNE experiment can further push the limit to
$|\Delta m^2_{31} - \Delta \overline m^2_{31}| < 8.1 \times 10^{-5} \mbox{eV}^2$
\cite{Barenboim:2017ewj}. In addition to causing difference in the oscillation
patterns for neutrinos and antineutrinos, the presence of CPT violation has
many other phenomenological consequences, such as
neutrino-to-antineutrino transitions \cite{Diaz:2016fqd} and baryogenesis
\cite{CPT-Baryon}. 

Possible violation of the CPT theorem can arise from
Lorentz violation \cite{LorentzViolation,Greenberg:2002uu}, non-locality
\cite{Barenboim:2002tz}, non-commutative geometry \cite{NonCommutative},
or Ether potential \cite{DeGouvea:2002xp}.
In this letter we provide a natural realization of CPT violation as
environmental dark NSI. Without introducing CPT symmetry breaking
at the Lagrangian level, a splitting in the neutrino and anti-neutrino
masses can arise when neutrinos travel through the DM medium. The Lorentz
and consequently CPT invariances are violated by the environmental
DM medium. Combining different types of neutrino oscillation experiments
can help us to identify this CPT violation.

{\it The Dark NSI} --
Neutrino oscillation can happen if neutrino masses are non-degenerate and
the mixing from flavor to mass eigenstates \cite{Pontecorvo,MNS} is nontrivial.
In vacuum, the neutrino oscillation is totally determined by the neutrino
mass matrix. However, the oscillation pattern can receive environmental effect
if neutrinos propagate through matter \cite{Wolfenstein:1977ue, resonant}.
From the forward scattering with matter particles, either electron or nuclei,
neutrino propagator can receive corrections \cite{MatterEffect, Mohapatra:1998rq}.
Even without mass term, neutrino oscillation can happen in matter
\cite{Wolfenstein:1977ue}. 

If DM is a fundamental particle, our universe is immersed in a sea
of DM particles. According to the 
astrophysical constraints, the local
DM energy density is $\rho_\chi \approx 0.47\,\mbox{GeV/cm}^3$
\cite{Catena:2009mf} and its number density is inversely proportional to
its mass $n_\chi = \rho_\chi / m_\chi$. With small enough mass, there would be a
plenty of DM particles surrounding us. Due to the Pauli exclusion principle,
the light DM ($\lesssim 100$~eV) can only be bosons, either scalar or vector particles.
In this letter, we first focus on the scalar case while the conclusion can also
apply to the vector one. If the scalar DM particle
has interaction with neutrinos, the relevant Lagrangian is
\begin{equation}
- \mathcal L
=
  \frac 1 2 m^2_\phi \phi^2
+ \frac 1 2 M_{\alpha \beta} \bar \nu_\alpha \nu_\beta
+ y_{\alpha \beta} \phi \bar \nu_\alpha \nu_\beta
+ h.c. \,,
\label{eq:L}
\end{equation}
with a Yukawa coupling between the light DM $\phi$ ($\equiv \chi$)
and neutrinos.
\begin{figure}[t]
\centering
\includegraphics[width=0.5\textwidth]{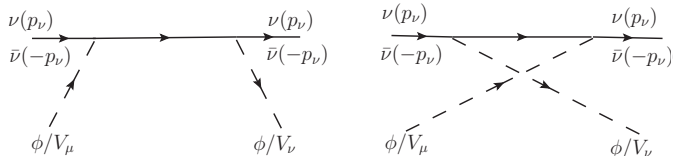}
\caption{The neutrino forward scattering with scalar $\phi$ or vector ($V$)
         light DM particles.}
\label{fig:forward}
\end{figure}
 
When neutrinos propagate through the DM medium, both neutrino and DM 
particles are present as real particles. The forward scattering in
\gfig{fig:forward} is then described by the scattering matrix element
\begin{equation}
  \langle \nu_\alpha (p_\nu) \phi(p_\phi)
| T e^{i \int d^4 x \mathcal L} |
  \nu_\beta (p_\nu) \phi (p_\phi) \rangle \,,
\label{eq:Smatrix}
\end{equation}
with neutrino and DM being the external particles. By definition,
the forward scattering has zero momentum exchange among the external particles.
Consequently, the initial- and final-state neutrinos (DM particles)
have exactly the same momentum $p_\nu$ ($p_\phi$). The resulting correction to
the two-point function can generally decompose as
$\delta \Gamma \equiv \delta \Gamma^\mu \gamma_\mu + \delta M$ and the neutrino
(anti-neutrino) Hamiltonian expands into
\begin{equation}
  H
\approx
  \frac {(M + \delta M) (M + \delta M)^\dagger}{2 E_\nu}
\mp \delta \Gamma^0 \,.
\label{eq:H}
\end{equation}
While the matter potential $\delta \Gamma^0$ receives an opposite sign, the
mass term correction $\delta M$ is the same for both neutrino and antineutrino modes.
The effective Hamiltonian \geqn{eq:H} can generally apply for various
matter effects that neutrino can experience \cite{MatterEffect,scalarNSI}.
Considering the fact that
the DM particles around the Earth are non-relativistic, we
just need to keep the dominant time component, $\fp_\phi \approx m_\phi \gamma_0$.
The leading-order contribution is
\begin{equation}
  \delta \Gamma_{\alpha \beta}
\approx
  \sum_j y_{\alpha j} y^*_{j \beta}
  \frac {\rho_\chi}{m^2_\phi E_\nu}
  \gamma_0 \,.
\label{eq:dGamma}
\end{equation}

An interesting feature is that Eq.~\geqn{eq:dGamma} has energy dependence, rather
than the energy-independent SM matter potential \cite{MatterEffect} or
the mass term correction from the scalar NSI \cite{scalarNSI}.
This leads to significantly
different phenomenological consequences. Since the first term in \geqn{eq:H}
is also inversely proportional to the neutrino energy $E_\nu$, the correction from
\geqn{eq:dGamma} then appears as correction to the mass squared term instead,
\begin{equation}
  H
=
  \frac {M^2}{2 E_\nu}
\mp
  \frac 1 {E_\nu}
  \sum_j y_{\alpha j} y^*_{j \beta} \frac {\rho_\chi}{m^2_\phi}
\equiv
  \frac {M^2 \pm \delta M^2}{2 E_\nu} \,,
\label{eq:HM}
\end{equation}
where $\delta M^2_{\alpha \beta} \equiv \mp
\frac {2 \rho_\chi}{m^2_\phi} \sum_j y_{\alpha j} y^*_{j \beta}$.
From its coupling with DM, neutrinos receive an opposite mass
squared correction from that of anti-neutrinos. This is essentially an apparent
violation of the CPT symmetry due to the environmental effect.

At first sight, it may seem strange why a chirality-flipping Yukawa coupling
in \geqn{eq:L} can lead to chirality-conserving correction \geqn{eq:dGamma}.
Although it is true that Yukawa coupling does flip chirality, two Yukawa
vertices in \gfig{fig:forward} can flip the neutrino chirality twice and
conserve the neutrino chirality. In addition, the non-zero momentum flow
in the neutrino propagator of \geqn{eq:L} provides $1/E_\nu$ dependence
and promotes the $\gamma_0$ term to correction of the neutrino mass squared term.

The earlier studies \cite{fuzzyDM} focused on the fuzzy DM scenario
which is equivalent to replacing the scalar DM field in 
\geqn{eq:L} by $\phi \rightarrow \sqrt{2 \rho_\chi} \cos (m_\phi t) / m_\phi$
with time variation. Nevertheless,
this effect is essentially correction to the neutrino mass $\delta M$
rather than the mass squared term, $\delta M^2$. As already indicated in
\geqn{eq:H}, the correction to the neutrino mass term has no
sign difference between neutrino and anti-neutrino. For a complex scalar,
$\phi = |\phi| e^{i m_\phi t}$,
the $\delta M^2$ correction is time independent and then
the time-dependent $\delta M$ term in \geqn{eq:H} can be safely ignored if
we only consider the time-averaged data.

Note that being fuzzy DM is not necessary for sizable dark NSI effect
on neutrino oscillation. With proportionally larger Yukawa coupling and mass,
the light DM can have large enough dark NSI as the fuzzy one.
For example, the effect scales as $y_\phi / m_\phi$ in the
condensation case. It is definitely possible to relax the mass and Yukawa coupling
range while maintaining the size of dark NSI. The forward scattering contribution
is actually of the same order as the condensation one.
While the former scales as $y^2_\phi \rho_\chi / m^2_\phi$ and contributes to
$\delta M^2$, the later scales as $y_\phi \sqrt{\rho_\chi} / m_\phi$ and
contributes to $\delta M$.

In addition, \cite{NuDamping} studied the matter effect from both
fermion and scalar fields. Their study is for totally different
environment, in supernova or the host plasma of the Early Universe. With a
$\bar f_R \nu_L \phi$ term, the neutrino can receive matter potential from
both $f$ and $\phi$ backgrounds that are present in supernova or the Early
Universe. The fermion $f$ can be either a DM fermion or sterile neutrino.
For both cases, the matter effect is always recognized as potential, rather
than correction to the neutrino mass squared term.

\begin{figure}[t]
\centering
\includegraphics[height=0.4\textwidth,width=4.75cm,angle=-90]{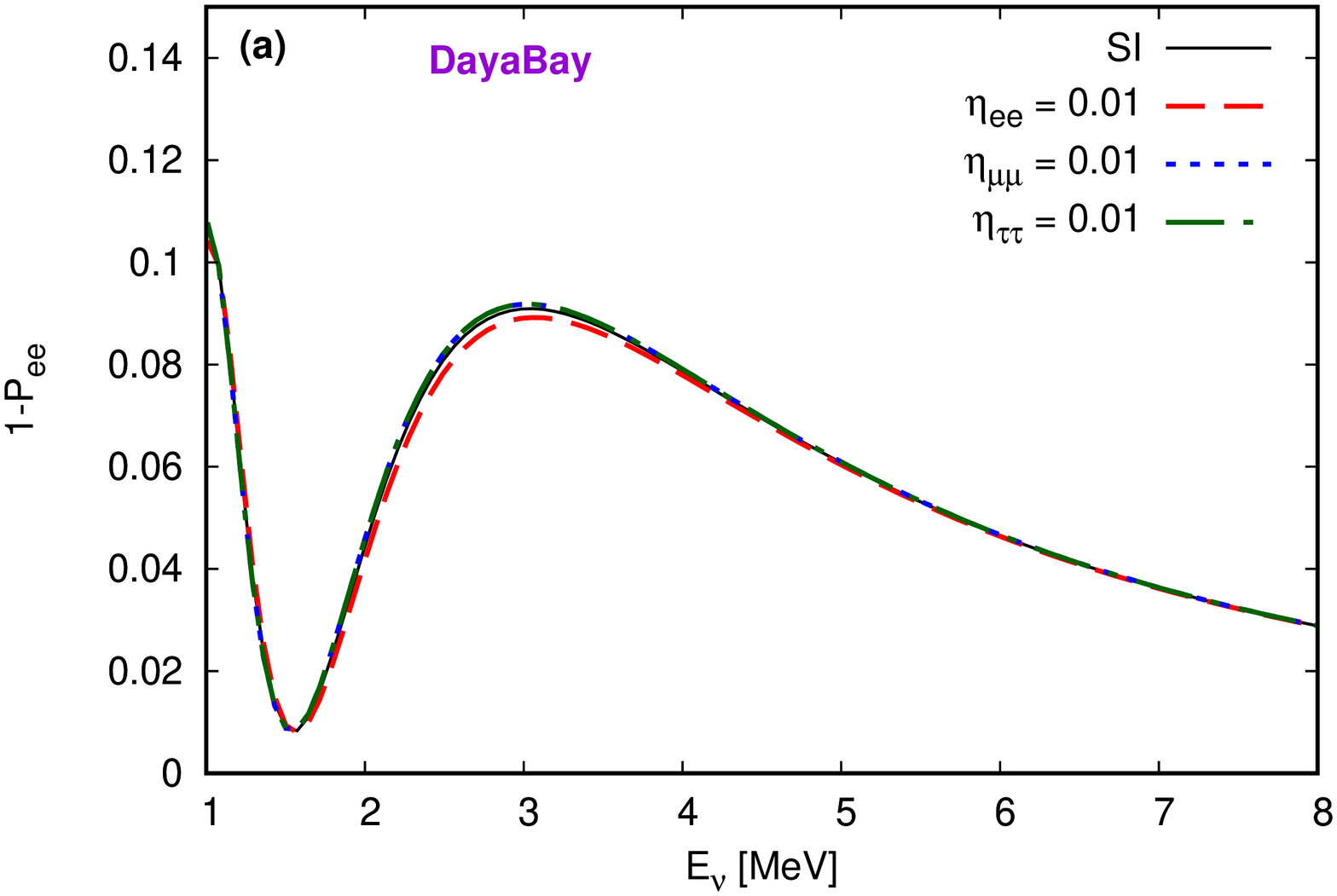}
\includegraphics[height=0.4\textwidth,width=4.75cm,angle=-90]{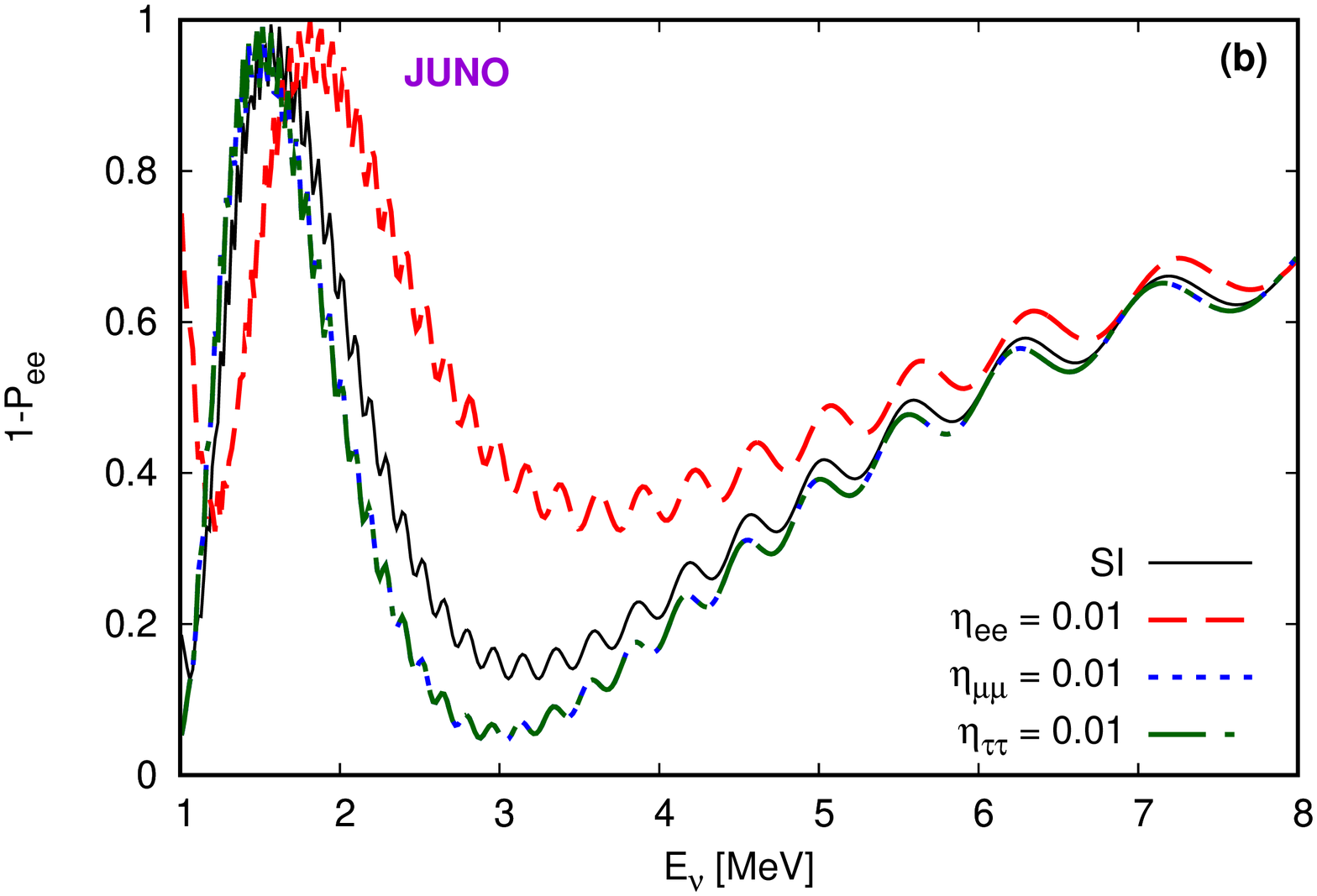}
\caption{The effect of the dark Non-Standard Interactions (NSI) in Eq.~(\ref{eq:darkNSI}) on (a) short-baseline neutrino oscillation
at Daya Bay and (b) medium-baseline neutrino oscillation at JUNO.  SI refers to Standard Interactions.}
\label{fig:reactor}
\end{figure}

{\it Phenomenological Consequences} --
To get a better sense of the dark NSI, we 
parametrize the correction to $M M^\dagger$ in general as
\begin{equation}
  \delta M^2
\equiv
  \Delta m^2_a
\left\lgroup
\begin{matrix}
  \eta_{ee} & \eta_{e \mu} & \eta_{e \tau} \\
  \eta^*_{e \mu} & \eta_{\mu \mu} & \eta_{\mu \tau} \\
  \eta^*_{e \tau} & \eta^*_{\mu \tau} & \eta_{\tau \tau}
\end{matrix}
\right\rgroup \,,
\label{eq:darkNSI}
\end{equation}
where the atmospheric mass squared difference
$\Delta m^2_a \equiv \Delta m^2_{31} \approx 2.7 \times 10^{-3} \mbox{eV}^2$
is the larger one of the two characteristic scales in $M M^\dagger$ while
dimensionless parameters $\eta_{\alpha \beta}$ parameterize the
size of the dark NSI in the unit of $\Delta m^2_a$. All simulations
are done with NuPro \cite{NuPro}. Sizable effect appears with
$\delta M^2_{\alpha \beta} \sim \Delta m^2_{ij}$, or equivalently,
$m_\phi / y_{\alpha j} \sim \sqrt{2 \rho_\chi / \Delta m^2_{ij}}
\approx (0.043 \sim 0.25)\,\mbox{eV}$ for $\rho_\chi = 0.3\,\mbox{GeV}/\mbox{cm}^3$,
$\Delta m^2_{21} = 7.55 \times 10^{-5}\,\mbox{eV}^2$ and
$\Delta m^2_{31} = 2.50 \times 10^{-3}\,\mbox{eV}^2$ \cite{nuGlobal}.
\begin{figure}[t]
\centering
\includegraphics[height=0.4\textwidth,width=4.75cm,angle=-90]{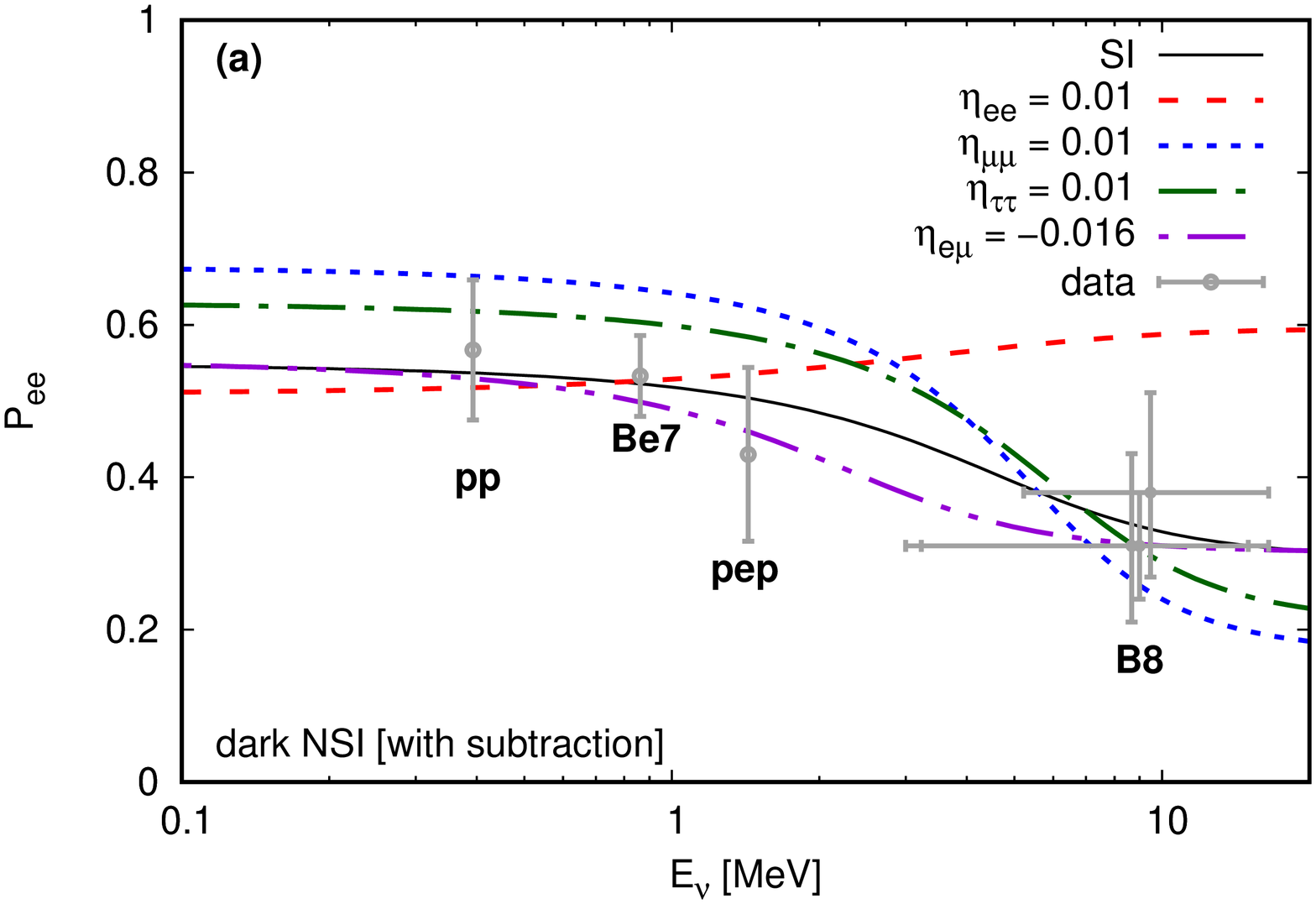}
\includegraphics[height=0.37\textwidth,width=4.75cm,angle=-90]{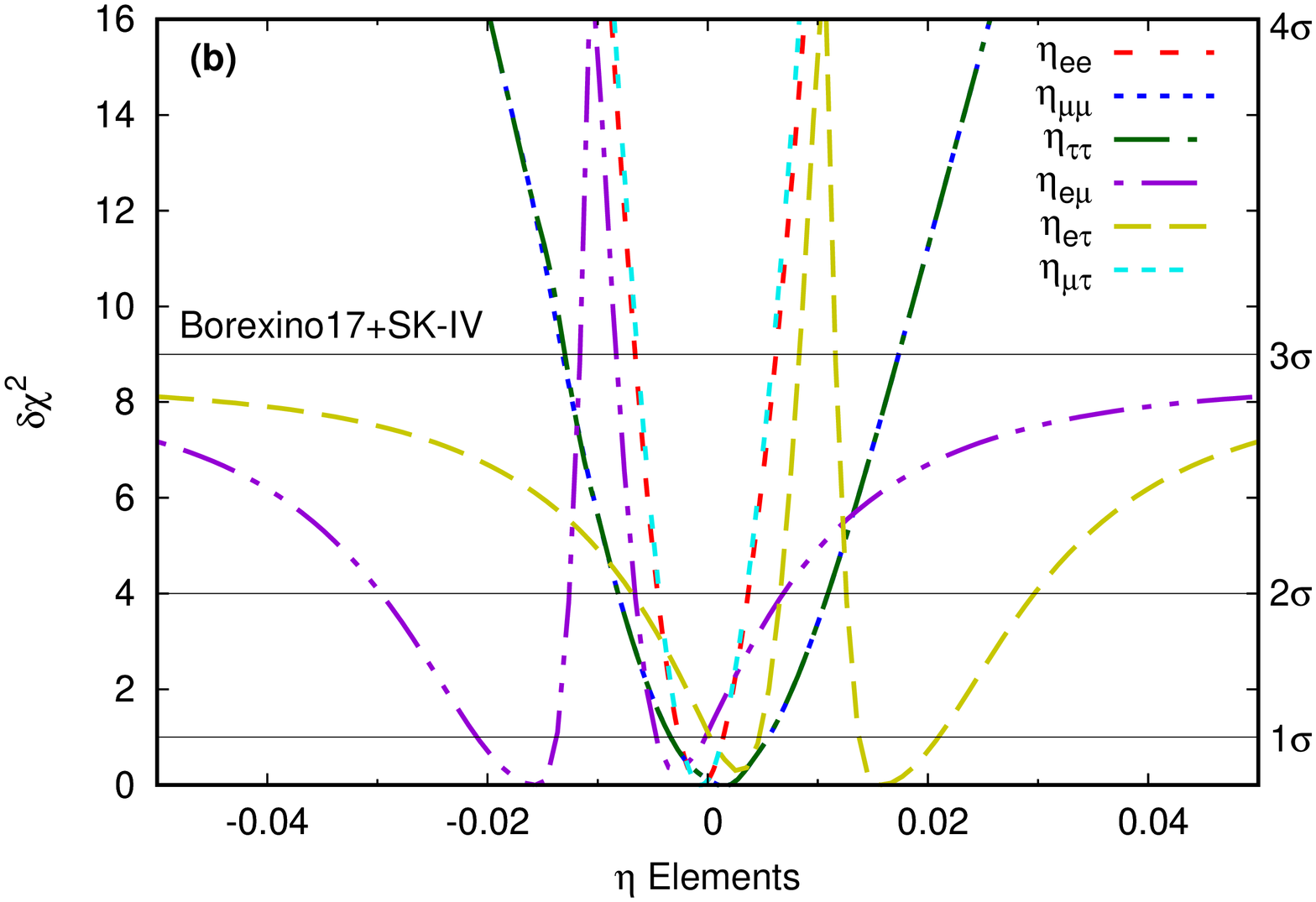}
\caption{(a) The dark NSI effect on solar neutrino transition probability
and (b) the $\chi^2$ fit to the Borexino \cite{Borexino17} and SK-IV \cite{SK} data.}
\label{fig:solar}
\end{figure}
The previous studies \cite{fuzzyDM} have used the time variation of
experimental data to constrain the CPT conserving part $\delta M \propto y/m_\phi$
in \geqn{eq:H} to roughly $10\sim15\%$ uncertainty of the mass
scale. Since the CPT violating correction $\delta M^2 \sim \mathcal O(\delta M)^2$,
we can expect percentage level of CPT violation and can be even as large as
$\eta_{\alpha \beta} \sim \mathcal O(0.1)$ for $2\sim3 \sigma$ confidence level.
Note that only the ratio $m_\phi / y$ matters and the dark matter mass can span
a large range, $(10^{-22} \sim 10^{-5})\,\mbox{eV}$, as long as the coupling
scales proportionally within the perturbative range.

Being a correction to the mass squared term, the dark NSI effect is
energy independent according to \geqn{eq:HM}. Even at low
energy, the dark NSI effect can be significant, for example, in the
solar and reactor neutrino oscillations. Most importantly, the neutrino
and anti-neutrino modes have the opposite signs which provide an extra
way of identification from the scalar NSI \cite{scalarNSI}.

In \gfig{fig:reactor} we show the dark NSI effect on the reactor neutrino
oscillations. The effect at the Daya Bay experiment \cite{DayaBay} is quite
moderate since its oscillation is modulated by the larger mass squared difference
$\Delta m^2_{31}$. With $\eta_{\alpha \alpha} = 0.01$, the dark NSI
contributes only 1\% of $\Delta m^2_{31}$ which is just
around the Daya Bay precision. However, the effect is significant
at the medium-baseline JUNO experiment \cite{JUNO}. The lower-frequency oscillation
modulated by the smaller $\Delta m^2_{21}$ is just 3\% of $\Delta m^2_{31}$
and is comparable to the dark NSI.  The JUNO experiment can
significantly improve the probe of the dark NSI.

\begin{figure}[t]
\centering
\includegraphics[height=0.4\textwidth,width=4.75cm,angle=-90]{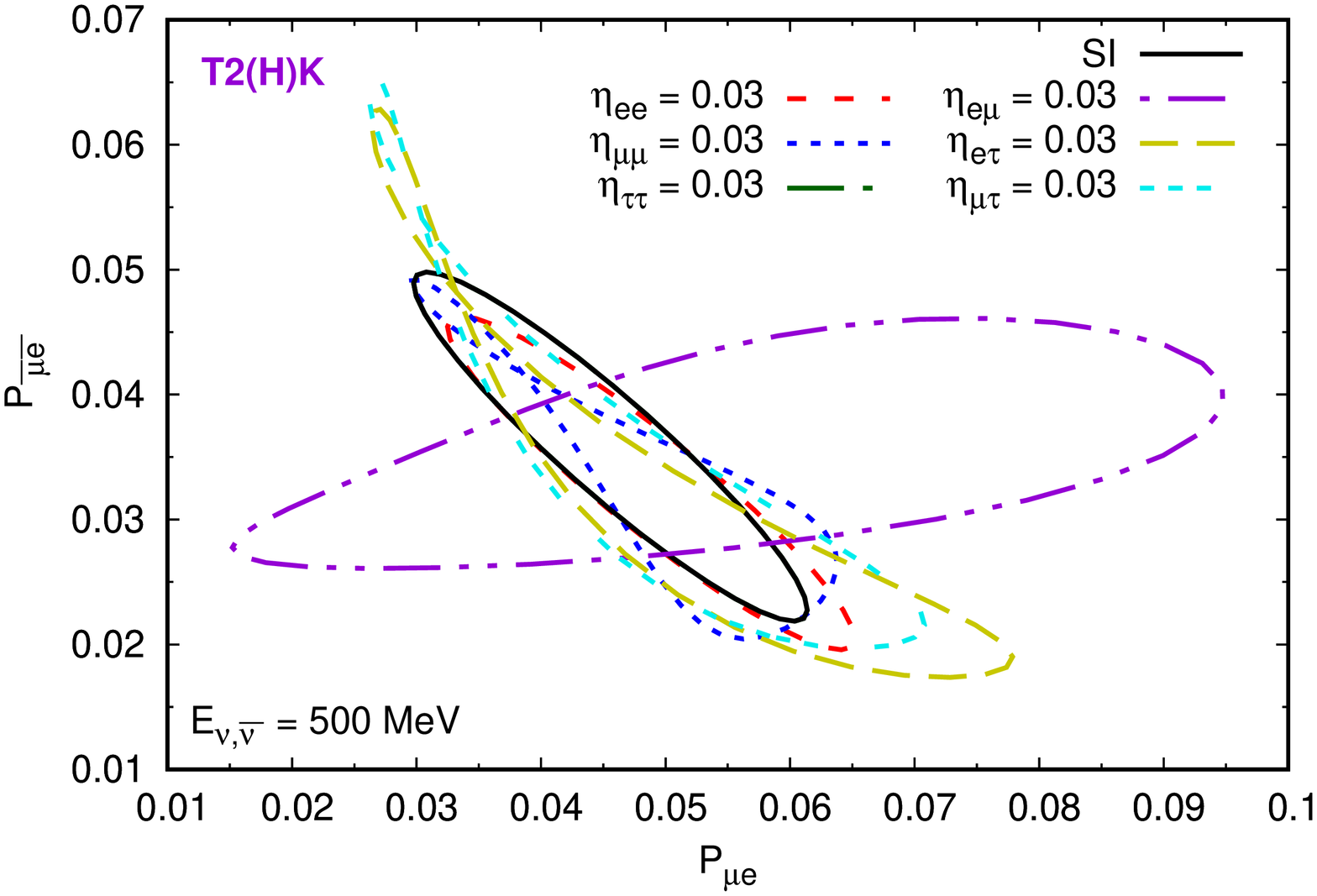}
\includegraphics[height=0.4\textwidth,width=4.75cm,angle=-90]{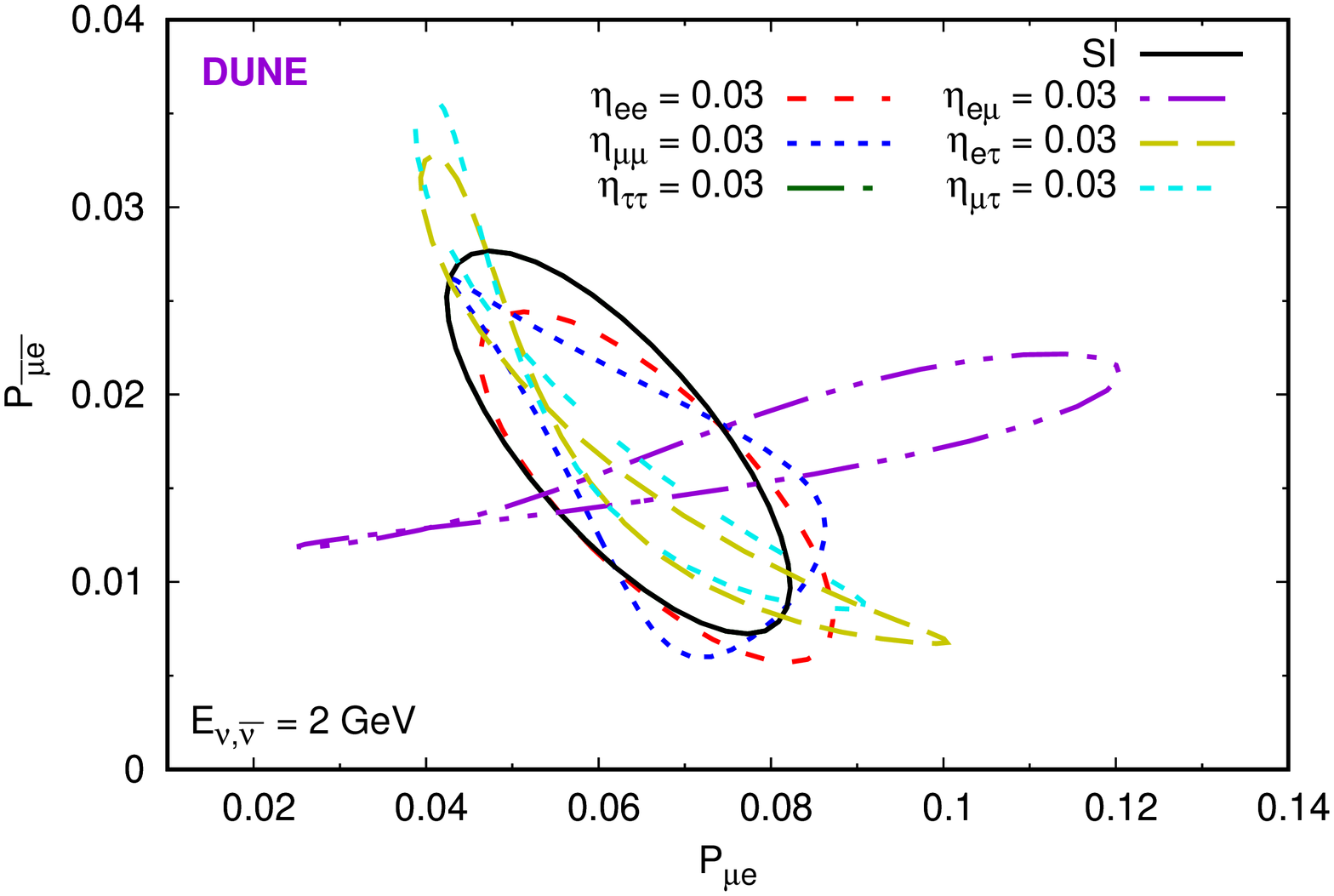}
\caption{The dark NSI effect at T2(H)K \cite{T2K} and DUNE \cite{DUNE}.}
\label{fig:accelerator}
\end{figure}
For the 1-2 mixing sector, the KamLAND reactor anti-neutrino measurement
\cite{KamLAND} has mismatched contour from the solar neutrino measurements
at SNO \cite{SNO}, Borexino \cite{Borexino17}, and SK \cite{SK} for both
the mass square difference $\Delta m^2_s \equiv \Delta m^2_{21}$ and the
solar mixing angle $\theta_s \equiv \theta_{12}$ \cite{PDG18,nuGlobal}.
While KamLAND gives
$\Delta m^2_{21} = 7.54^{+0.19}_{-0.18} \times 10^{-5} \, \mbox{eV}^2$
and $\sin^2 \theta_{12} = 0.316^{+0.034}_{-0.026}$,
the solar data prefers
$\Delta m^2_{21} = 4.82^{+1.20}_{-0.60} \times 10^{-5} \, \mbox{eV}^2$
and $\sin^2 \theta_{12} = 0.310 \pm 0.014$ \cite{SK}.
It is possible for the dark NSI to reconcile these two datasets.
Fitting the Borexino 2017 \cite{Borexino17} and the SK \cite{SK} data
sets, the $\delta \chi^2$ curves in \gfig{fig:solar} for the $\eta_{e\mu}$ and
$\eta_{e\tau}$ elements clearly shows an extra minimum which is even
lower than the minimum with vanishing dark NSI.
The coupling of bosonic DM with neutrino provides a natural
realization of the CPT violation to explain the long-standing
discrepancy. Between the two local minima, there is a high peak around
$\eta_{e\mu} \approx -0.01$.

We use the 2-$\nu$ formalism
\begin{equation*}
  M^2_{2\nu, 2\bar\nu}
\equiv
  \Delta m^2_s
\left\lgroup
\begin{matrix}
  s^2_s & c_s s_s \\
  c_s s_s & c^2_s
\end{matrix}
\right\rgroup
\pm
  c_{\nu, \bar \nu}
  \Delta m^2_a
\left\lgroup
\begin{matrix}
  \eta_{ee} & \eta_{e \mu} \\
  \eta^*_{e \mu} & \eta_{\mu \mu}
\end{matrix}
\right\rgroup \,,
\end{equation*}
where $c_{\nu, \bar \nu} = 1$, to quantitatively understand these results.
Diagonalizing $M^2_{2 \nu}$ and $M^2_{2 \bar \nu}$
gives two sets of $(\Delta m^2_s)^{\nu, \bar \nu}_{\rm eff}$ and
$(\theta_s)^{\nu, \bar \nu}_{\rm eff}$ to account for the different
measured values from the reactor anti-neutrino and solar neutrino experiments. 

Since the experimentally measured variables are those effective ones of the
$M^2_{2 \nu}$ and $M^2_{2 \bar \nu}$ for the neutrino and anti-neutrino
modes, respectively, it is more convenient to use the subtraction trick
\cite{scalarNSI}. In other words, we first reconstruct $M^2_{2 \bar \nu}$
with the measured or effective variables. Correspondingly, $c_\nu = 2$
and $c_{\bar \nu} = 0$. As $c_s s_s \approx 2/3$ and
$\Delta m^2_a / \Delta m^2_s \approx 30$, the off-diagonal elements of
$M^2_{2 \nu}$ vanishes with $\eta_{e\mu} \approx -0.01$, leading to unrealistic
$(\theta_s)^\nu_{\rm eff} = 0$ and hence the high peak in the $\delta \chi^2$ curve.

\begin{figure}[t]
\centering
\includegraphics[height=0.4\textwidth,width=4.75cm,angle=-90]{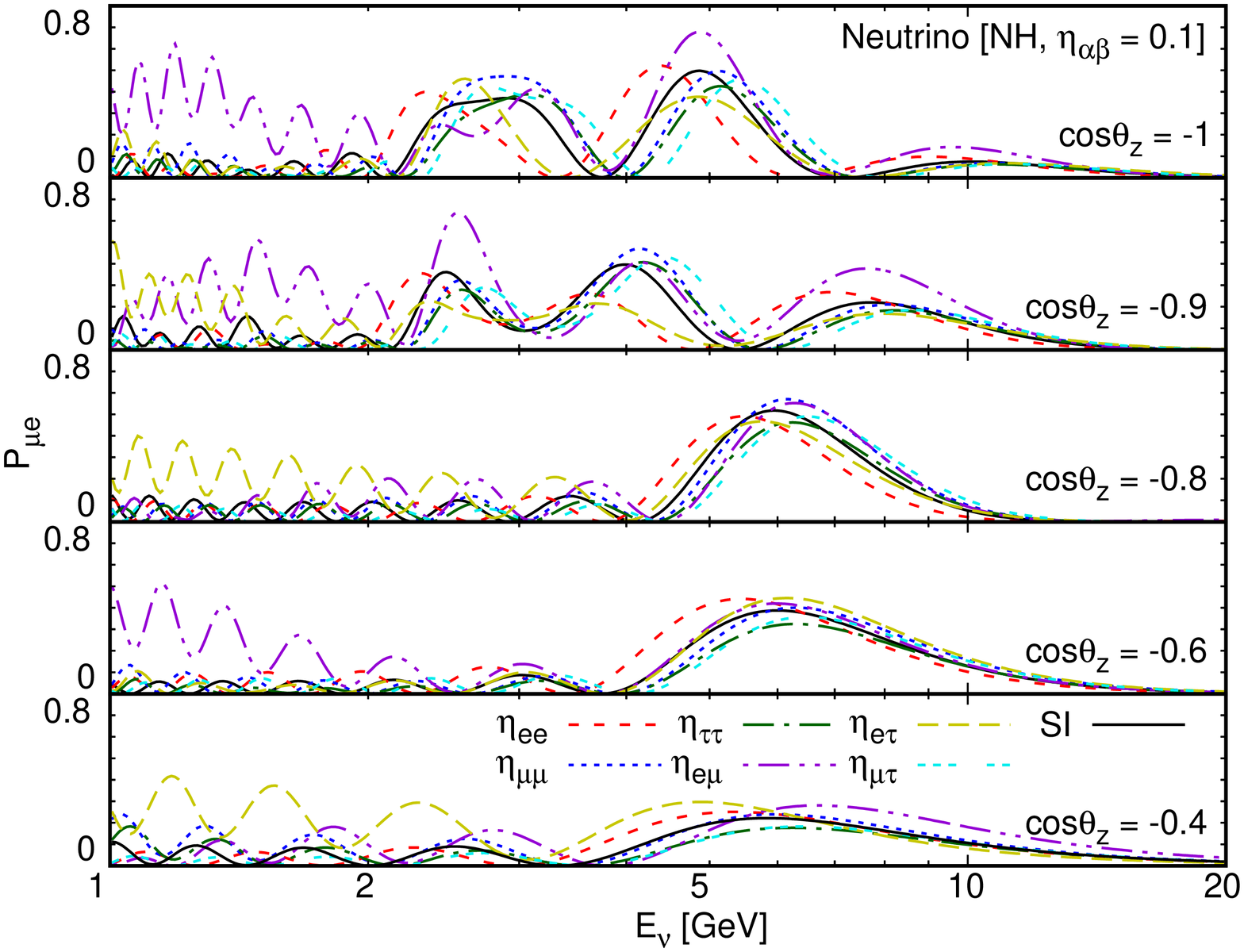}
\includegraphics[height=0.4\textwidth,width=4.75cm,angle=-90]{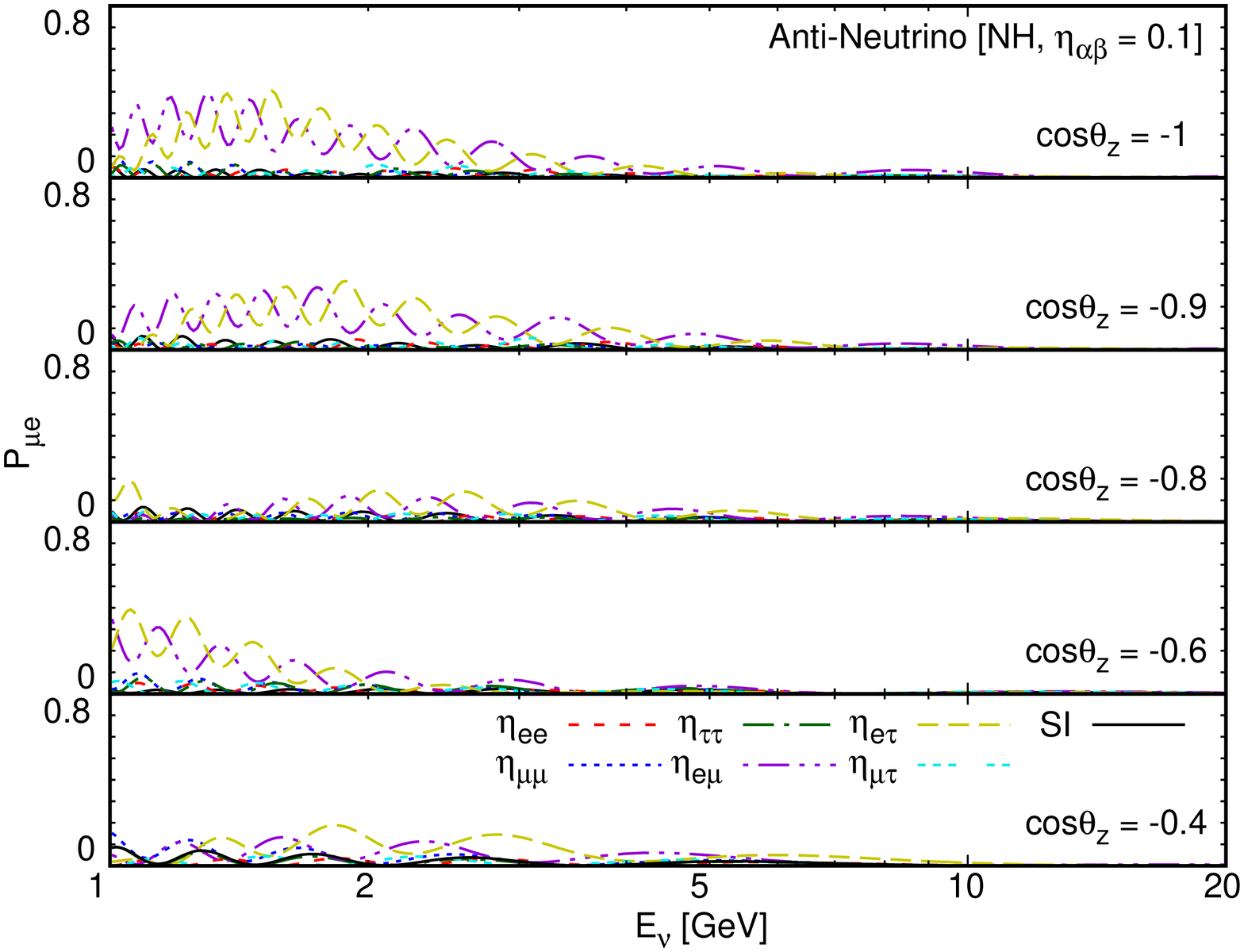}
\caption{The dark NSI effect on the atmospheric neutrino oscillation probabilities
         for various zenith angle $\theta_z$, simulated
         with the algorithm in \cite{atmos}.}
\label{fig:atmos}
\end{figure}
The dark NSI has sizable effect at both
low and high energy regions, crossing the black SI curve in the intermediate region.
The $\delta \chi^2$ curve has a global minimum at
$\eta_{e \mu} = - \eta_{\mu e} \approx -0.16$
which approaches the smaller $\Delta m^2_{21}$ solution. At $2\sigma$
level, $\epsilon_{e\mu}$ and $\epsilon_{e\tau}$ can be as large as $\pm 0.03$
which can further relax to $\pm 0.1$ at $3\sigma$ confidence level.
Better measurement of the solar neutrino fluxes at the SNO+ \cite{SNO+}
and Jinping neutrino \cite{JinPing} experiments can help
to identify the dark NSI.

With $\eta_{\alpha \beta} = 3\%$, the dark NSI effect on the CP
measurement with accelerator neutrinos is already significant, see
\gfig{fig:accelerator}. While most of the $\eta_{\alpha \beta}$ elements
deforms the biprobability contour around the SI one, the deviation by
$\eta_{e \mu}$ can significantly change the picture. This is understandable
since the CP measurement is mainly about the $\mu \rightarrow e$ transitions
and hence is sensitive to any modification in the $e \mu$ element.
Unfortunately, the $\eta_{e \mu}$ element at percentage level
is not well constrained by the
solar neutrino oscillation, see \gfig{fig:solar}. Other complementary
searches are necessary to guarantee the CP sensitivity at the accelerator
type experiments against the dark NSI.

The atmospheric neutrino oscillation might provide such a complementary
channel. As shown in \gfig{fig:atmos}, the $\eta_{e \mu}$ element
can significantly modify the oscillation behaviors, especially
around the MSW resonances which is the most important region to the
neutrino mass hierarchy measurement with atmospheric neutrinos.
With large event rate at PINGU \cite{PINGU} and
ORCA \cite{ORCA}, or the ability of INO \cite{INO} in distinguishing
neutrino from anti-neutrino, good sensitivity on
$\eta_{e \mu}$ can be expected with prior knowledge on
the neutrino mass hierarchy.

{\it Conclusion} --
The CPT violation can appear without breaking the CPT symmetry in the
fundamental Lagrangian. Instead, it can arise as environmental effect
and act as a manifestation of the coupling between neutrino and light DM.
A new channel of probing the light DM appears in the neutrino oscillation.
In addition to affecting the low-energy reactor and solar
neutrino oscillations, the dark NSI can phenomenologically fake the genuine
Dirac CP phase in the accelerator experiments. To guarantee the CP sensitivity,
a synergy among various types of neutrino experiments is necessary.

\section*{Acknowledgements}

The work of SFG was supported in part by JSPS KAKENHI Grant Number JP18K13536.  H.M. was supported in part by the NSF grant PHY-1638509, by the U.S. DOE Contract DE-AC02-05CH11231, by the JSPS Grant-in-Aid for Scientific Research (C) (17K05409), and MEXT Grant-in-Aid for Scientific Research on Innovative Areas (15H05887, 15K21733).  Both SFG and HM were supported in part by World Premier International (WPI) Research Center Initiative, MEXT, Japan.
SFG is grateful to the NPC Fellowship Award and the hospitality provided
by the Neutrino Physics Center of Fermilab where this work was initiatied.

\newpage

\setcounter{equation}{0}
\renewcommand{\theequation}{S\arabic{equation}}

\begin{center}
{\bf Supplementary Material}
\end{center}

When neutrinos propagate through dark matter, both neutrino and dark matter
particles are present as real particles. The forward scattering is then 
described by scattering matrix element
\begin{equation}
  \langle \nu_\alpha (p_\nu) \phi(p_\phi)
| T e^{i \int d^4 x \mathcal L} |
  \nu_\beta (p_\nu) \phi (p_\phi) \rangle \,.
\label{eqS:Smatrix}
\end{equation}
By definition, forward scattering has no momentum exchange among particles.
Consequently, the initial- and final-state neutrinos (dark matter particles)
have exactly the same momentum $p_\nu$ ($p_\phi$). A direct consequence is
that the two scalar dark matter fields $\phi$ share exactly the same wave
function. The scalar dark matter particles can be replaced as
$\phi \rightarrow \sqrt{2 \rho_\chi} / m_\phi$ times a creation
or annihilation operator for the initial- or final-state states,
$|\phi \rangle \rightarrow (\sqrt{2 \rho_\chi} / m_\phi) a^\dagger_\phi |0\rangle$
or $\langle \phi | \rightarrow (\sqrt{2 \rho_\chi} / m_\phi) \langle 0 | a_\phi$,
respectively. To eliminate $a^\dagger_\phi$ and $a_\phi$, the leading
contribution is the one shown in \gfig{fig:forward} with two $\phi \bar \nu \nu$
vertices,
\begin{equation}
 i \delta S_{\beta \alpha}
\equiv
  \frac 1 2
\langle \nu_\beta \phi | T
  (i y_{ij} \phi \bar \nu_i \nu_j)
  (- i y^*_{kl} \phi^* \bar \nu_k \nu_l)
| \nu_\alpha \phi \rangle \,,
\label{eq:scalarDM-V1}
\end{equation}
for the transition $\nu_\alpha \rightarrow \nu_\beta$.
With second quantization, the neutrino field and state are defined as
$\nu = a u + b^\dagger v$ and $|\nu \rangle = a^\dagger |0\rangle$.
Consequently, contraction can only happens between $\nu$ and $|\nu\rangle$
as well as between $\bar \nu$ and $\langle \nu |$. Since there is no difference
between the two vertices in \geqn{eq:scalarDM-V1}, we can use the contraction
of neutrino operators to fix the order of these two vertices, $\langle \nu|$
contracts with the first and $|\nu \rangle$ with the second, as shown in
\geqn{eq:scalarDM-nu}. Then there are two different
ways of contracting the DM field $\phi$ and its external state $|\phi\rangle$,
\begin{eqnarray}
 i \delta S_{\beta \alpha}
& = &
  \frac 1 2
  \bcontraction[0.5ex]{\langle} {\nu} {_\beta \phi | (i y_{ij} \phi} {\bar \nu}
  \bcontraction[0.5ex]{\langle \nu_\beta  \phi | (i y_{ij} \phi \bar \nu_i} {\nu}{_j) (- i y^*_{kl} \phi} {^*\bar \nu}
  \bcontraction[0.5ex]{\langle \nu_\beta  \phi | (i y_{ij} \phi \bar \nu_i \nu_j) (- i y^*_{kl} \phi^* \bar \nu_k} {\nu}{_l) |} {\nu}
  \acontraction[0.5ex]{\langle \nu} {_\beta \phi} {| (i y_{ij}} {\phi}
  \acontraction[0.5ex]{\langle \nu_\beta  \phi | (i y_{ij} \phi \bar \nu_i \nu_j) (- i y^*_{kl}} {\phi} {^*\bar \nu_k \nu_l) | \nu} {_\alpha \phi}
  \langle \nu_\beta  \phi | (i y_{ij} \phi \bar \nu_i \nu_j) (- i y^*_{kl} \phi^* \bar \nu_k \nu_l) | \nu_\alpha \phi \rangle
\nonumber
\\[1mm]
& + &
  \frac 1 2
  \bcontraction[0.5ex]{\langle} {\nu} {_\beta \phi | (i y_{ij} \phi} {\bar \nu}
  \bcontraction[0.5ex]{\langle \nu_\beta  \phi | (i y_{ij} \phi \bar \nu_i} {\nu}{_j) (- i y^*_{kl} \phi} {^*\bar \nu}
  \bcontraction[0.5ex]{\langle \nu_\beta  \phi | (i y_{ij} \phi \bar \nu_i \nu_j) (- i y^*_{kl} \phi^* \bar \nu_k} {\nu}{_l) |} {\nu}
  \acontraction[0.5ex]{\langle \nu} {_\beta \phi} {| (i y_{ij} \phi \bar \nu_i \nu_j) (- i y^*_{kl}} {\phi}
  \acontraction[1.0ex]{\langle \nu_\beta  \phi | (i y_{ij}} {\phi} {\bar \nu_i \nu_j) (- i y^*_{kl} \phi^* \bar \nu_k \nu_l) | \nu} {_\alpha \phi}
  \langle \nu_\beta  \phi | (i y_{ij} \phi \bar \nu_i \nu_j) (- i y^*_{kl} \phi^* \bar \nu_k \nu_l) | \nu_\alpha \phi \rangle \,.
\label{eq:scalarDM-nu}
\end{eqnarray}
The remaining one neutrino and one anti-neutrino fields would contract to
become a neutrino propagator, as depicted in \gfig{fig:forward}.
The sample procedure can be repeated for anti-neutrino 
$\bar \nu_\alpha \rightarrow \bar \nu_\beta$ transition,
\begin{eqnarray}
 i \delta \overline S_{\beta \alpha}
& = &
  \frac 1 2
  \bcontraction[1.5ex] {\langle} {\bar \nu} {_\beta  \phi | (i y_{ij} \phi \bar \nu_i \nu_j) (- i y^*_{kl} \phi^* \bar \nu_k} {\nu}
  \bcontraction[1.0ex] {\langle \bar \nu_\beta  \phi | (i y_{ij} \phi} {\bar \nu} {_i \nu_j) (- i y^*_{kl} \phi^* \bar \nu_k \nu_l) |} {\bar \nu}
  \bcontraction[0.5ex] {\langle \bar \nu_\beta  \phi | (i y_{ij} \phi \bar \nu_i} {\nu} {_j) (- i y^*_{kl} \phi^*} {\bar \nu}
  \acontraction[0.5ex] {\langle \bar \nu_\beta}{\phi}{| (i y_{ij}} {\phi}
  \acontraction[0.5ex] {\langle \bar \nu_\beta  \phi | (i y_{ij} \phi \bar \nu_i \nu_j) (- i y^*_{kl}} {\phi} {^* \bar \nu_k \nu_l) | \bar \nu_\alpha} {\phi}
  \langle \bar \nu_\beta  \phi | (i y_{ij} \phi \bar \nu_i \nu_j) (- i y^*_{kl} \phi^* \bar \nu_k \nu_l) | \bar \nu_\alpha \phi \rangle
\nonumber
\\[1mm]
& + &
  \frac 1 2
  \bcontraction[1.5ex] {\langle} {\bar \nu} {_\beta  \phi | (i y_{ij} \phi \bar \nu_i \nu_j) (- i y^*_{kl} \phi^* \bar \nu_k} {\nu}
  \bcontraction[1.0ex] {\langle \bar \nu_\beta  \phi | (i y_{ij} \phi} {\bar \nu} {_i \nu_j) (- i y^*_{kl} \phi^* \bar \nu_k \nu_l) |} {\bar \nu}
  \bcontraction[0.5ex] {\langle \bar \nu_\beta  \phi | (i y_{ij} \phi \bar \nu_i} {\nu} {_j) (- i y^*_{kl} \phi^*} {\bar \nu}
  \acontraction[0.5ex] {\langle \bar \nu_\beta}{\phi} {| (i y_{ij} \phi \bar \nu_i \nu_j) (- i y^*_{kl}} {\phi}
  \acontraction[1.0ex] {\langle \bar \nu_\beta  \phi | (i y_{ij}} {\phi} {\bar \nu_i \nu_j) (- i y^*_{kl} \phi^* \bar \nu_k \nu_l) | \bar \nu_\alpha} {\phi}
  \langle \bar \nu_\beta  \phi | (i y_{ij} \phi \bar \nu_i \nu_j) (- i y^*_{kl} \phi^* \bar \nu_k \nu_l) | \bar \nu_\alpha \phi \rangle
\label{eq:scalarDM-NU}
\end{eqnarray}

In the transition matrix $\delta S_{\beta \alpha}$ for neutrino propagation,
the contracted neutrino operators are already next to each other but for
the anti-neutrino one $\delta \overline S_{\beta \alpha}$, the neutrino
operators need odd number of permutations to put paired ones together.
This leads to a minus sign difference between the neutrino and anti-neutrino
cases
\begin{equation}
  \delta S_{\beta \alpha}
\equiv
  \bar u_\beta \delta \Gamma_{\beta \alpha} u_\alpha \,,
\quad
  \delta \overline S_{\beta \alpha}
\equiv
- \bar v_\alpha \delta \Gamma_{\alpha \beta} v_\beta \,.
\label{eq:vertex}
\end{equation}
Adding these corrections to the neutrino kinetic terms,
\begin{eqnarray}
&&
  \bar u_\beta (\fp_\nu - M + \delta \Gamma)_{\beta \alpha} u_\alpha a^\dagger a
\nonumber
\\
& = &
  \bar v_\alpha (- \fp_\nu - M + \delta \Gamma)_{\alpha \beta} v_\beta b b^\dagger
=
  0 \,.
\label{eq:kinetic}
\end{eqnarray}
From \geqn{eq:vertex} to \geqn{eq:kinetic}, the sign associated with
$\delta \Gamma$ is compensated by the permutation of neutrino operators
while a sign difference now appears in the momentum part. Or equivalently,
the effective propagator is the summation of all diagrams,
\begin{equation}
\hspace{-3mm}
  \frac i {\pm \fp_\nu - M}
  \sum^\infty_{n = 0}
\left(
  i \delta \Gamma
  \frac i {\pm \fp_\nu - M}
\right)^n
=
  \frac i {\pm \fp_\nu - M + \delta \Gamma} \,,
\label{eqS:propagator}
\end{equation}
for neutrino and anti-neutrino, respectively.

If we generally decompose the two-point function as
$\delta \Gamma \equiv \delta \Gamma_\mu \gamma^\mu + \delta M$, the neutrino
(anti-neutrino) Hamiltonian expands as
\begin{equation}
  H
\approx
  \frac {(M + \delta M) (M + \delta M)^\dagger}{2 E_\nu}
\mp \delta \Gamma_0 \,.
\label{eqS:H}
\end{equation}
While the matter potential $\delta \Gamma_0$ receives an opposite sign, the
mass term correction is the same for the neutrino and antineutrino modes.
The formalism \geqn{eq:kinetic} and \geqn{eqS:H} is quite general for various
matter effects that neutrino can experience \cite{MatterEffect,scalarNSI}.
Note that the neutrino (anti-neutrino) oscillation is described by $H$ ($H^T$),
respectively, due to the different flavor assignments in \geqn{eq:vertex}
and \geqn{eq:kinetic}.

The concrete form of the two-point functions $i \delta \Gamma_{\alpha \beta}$
can be written down according to the Feynman diagrams in \gfig{fig:forward},
\begin{eqnarray}
  \delta \Gamma_{\alpha \beta}
& = &
  \frac {\rho_\phi({\bf v}_\phi)}{m^2_\phi}
  \sum_j y_{\alpha j} y^*_{j \beta}
\\
& \times &
\left[
  \frac i {\pm (\fp_\nu + \fp_\phi) - m_\nu}
+ \frac i {\pm (\fp_\nu - \fp_\phi) - m_\nu}
\right] \,,
\nonumber
\end{eqnarray}
where $p^2_\nu = m^2_\nu$ for on-shell neutrinos. First, let us move the
$\gamma$ matrices to the numerator
\begin{eqnarray}
  \delta \Gamma_{\alpha \beta}
& = &
  \frac {i \rho_\phi({\bf v}_\phi)}{m^2_\phi}
  \sum_j y_{\alpha j} y^*_{j \beta}
\\
& \times &
\left[
  \frac {\pm (\fp_\nu + \fp_\phi) + m_\nu}
        {p^2_\phi + 2 p_\nu \cdot p_\phi}
+
  \frac {\pm (\fp_\nu - \fp_\phi) + m_\nu}
        {p^2_\phi - 2 p_\nu \cdot p_\phi}
\right] \,,
\nonumber
\end{eqnarray}
whether the denominators have been simplified as
$(p_\nu \pm p_\phi)^2 - m^2_\nu = p^2_\phi \pm 2 p_\nu \cdot p_\phi$
for on-shell neutrinos. Since the momentum of the non-relativistic
light DM is much smaller than the neutrino momentum,
$p_\phi \sim m_\phi (1, \vec v_\phi) \ll p_\nu$, the denominators
are dominated by $\pm 2 p_\nu \cdot p_\phi \approx \pm 2 m_\phi E_\nu$.
Then the common term
$\pm \fp_\nu + m_\nu$ in the two numerators cancel with each other,
leaving only the $\fp_\phi$ term. Considering the fact that
dark matter particles around Earth are quite non-relativistic nowadays, we
just need to keep the dominating time component, $\fp_\phi \approx m_\phi \gamma_0$.
In addition, from neutrino to anti-neutrino, the momentum in the propagator receive
a minus sign to account for the opposite fermion flow, leading to the overall
sign in \geqn{eq:vertex}. Keeping only the leading order, we can get
\begin{equation}
  \delta \Gamma_{\alpha \beta}
\approx
  \sum_j y_{\alpha j} y^*_{j \beta}
  \frac {\rho_\chi}{m^2_\phi E_\nu}
  \gamma_0 \,,
\label{eqS:dGamma}
\end{equation}
with the total density $\rho_\chi$ from averaging over the DM velocity distribution,
$\int \rho_\phi({\bf v}_\phi) d {\bf v}_\phi = \rho_\chi$.

An interesting feature is \geqn{eqS:dGamma} has energy dependence, rather
than the energy-independent SM matter potential \cite{MatterEffect} or
mass term correction from scalar NSI \cite{scalarNSI}.
This leads to significantly
different phenomenological consequences. Since the first term in \geqn{eqS:H}
is also inversely proportional to neutrino energy, the correction from
\geqn{eqS:dGamma} then appears as correction to the mass squared term,
\begin{equation}
  H
=
  \frac {M^2}{2 E_\nu}
\mp
  \frac 1 {E_\nu}
  \sum_j y_{\alpha j} y^*_{j \beta} \frac {\rho_\chi}{m^2_\phi}
\equiv
  \frac {M^2 \pm \delta M^2}{2 E_\nu} \,,
\label{eqS:HM}
\end{equation}
where $\delta M^2_{\alpha \beta} \equiv \mp
\frac {2 \rho_\chi}{m^2_\phi} \sum_j y_{\alpha j} y^*_{j \beta}$.
If neutrino travels inside the ordinary matter, there is an extra
contribution from the matter potential induced by the SM charged currents.
Due to the presence of
light DM, neutrinos receive opposite mass squared correction than
anti-neutrinos. This is essentially a manifest violation of CPT
symmetry due to environmental effect.

At first sight, it may seem strange why a helicity-flipping Yukawa coupling
in \geqn{eq:L} can lead to helicity-conserving correction \geqn{eqS:dGamma}.
Although it is true that Yukawa coupling does flip helicity, two Yukawa
vertices in \gfig{fig:forward} can flip the neutrino helicity twice and
conserve the neutrino helicity. In addition, the non-zero momentum flow
in the neutrino propagator of \geqn{eq:L} provides $1/E_\nu$ dependence
and promotes the $\gamma_0$ term to correction of the neutrino mass squared term.

For vector DM particle $V$, it can couple with neutrino current
\begin{equation}
- \mathcal L
\ni
  \frac 1 2 m^2_V V_\mu V^\mu
+ \frac 1 2 M_{\alpha \beta} \bar \nu_\alpha \nu_\beta
+ g_{\alpha \beta} V_\mu \bar \nu_\alpha \gamma^\mu \nu_\beta \,.
\end{equation}
Following the same procedure of sandwiching action $S$ with external fields
and contracting particle creation versus annihilation operators in pair,
we can derive the effective two-point function
\begin{eqnarray}
  \delta \Gamma_{\alpha \beta}
& = &
- g_{\alpha j} g_{j \beta}
  \frac {\rho_V({\bf v}_V)}{m^2_V}
  \epsilon_\alpha(p'_V) \epsilon^*_\beta(p_V)
\label{eq:vectorDM-V1}
\\
& \times &
\left[
  \gamma^\alpha \frac {\fp_\nu + \fp_V + m_\nu} {m^2_V + 2 p_\nu \cdot p_V} \gamma^\beta
+ \gamma^\beta  \frac {\fp_\nu - \fp_V + m_\nu} {m^2_V - 2 p_\nu \cdot p_V} \gamma^\alpha
\right] \,.
\nonumber
\end{eqnarray}
Since the incoming and outgoing dark matter particles share the same momentum,
$p_V = p'_V$, the two polarization vectors are actually the same,
$\epsilon(p'_V) = \epsilon(p_V) \equiv \epsilon$. In addition, it is possible to
choose convention to make the polarization vectors real. Then the indices
$\alpha$ and $\beta$ in \geqn{eq:vectorDM-V1} can interchange with each other
and consequently we can first factorize out the two $\gamma$ matrices on the side,
\begin{eqnarray}
  \delta \Gamma_{\alpha \beta}
& = &
- g_{\alpha j} g_{j \beta}
  \frac {\rho_V({\bf v}_V)}{m^2_V}
  \epsilon_\alpha(p_V) \epsilon_\beta(p_V)
\\
& \times &
  \gamma^\alpha
\left[
  \frac {\fp_\nu + \fp_V + m_\nu} {m^2_V + 2 p_\nu \cdot p_V}
+ \frac {\fp_\nu - \fp_V + m_\nu} {m^2_V - 2 p_\nu \cdot p_V}
\right]
  \gamma^\beta \,.
\nonumber
\end{eqnarray}
Then, we can use the same argument as the scalar case to eliminate the
$\fp_\nu + m_\nu$ terms in the numerator and $\fp_V \approx m_V \gamma_0$.

With non-relativistic
dark matter, the three polarization vectors can be chosen as the three spatial
unit vector along $x$, $y$, and $z$ axes, $\epsilon^\mu_i = (0, {\bf e}_i)$,
respectively.
For $m_V \ll E_\nu$, we only need to consider the $\fp_V$ term. Since
DM is non-relativistic, its contribution is dominated by $\fp_V \approx m_V \gamma_0$.
The two identical polarization vectors $\epsilon_\alpha$ and $\epsilon_\beta$
can symmetrize their indices, $\epsilon_\alpha \epsilon_\beta = \epsilon_\beta \epsilon_\alpha$.
This significantly simplifies the $\gamma$ matrices,
$\epsilon_\alpha \epsilon_\beta \gamma^\alpha \gamma^0 \gamma^\beta
=
  2 (\epsilon \cdot \gamma) \epsilon^0
+ \gamma^0$.
Then the effective potential reduces to a form close to fermion propagator
with at most linear combination of $\gamma$ matrices,
Since the polarization vectors are orthogonal and have only spatial components,
the $2 (\epsilon \cdot \gamma) \epsilon^0$ term vanishes at the leading order.
The two-point function then simplifies to
\begin{equation}
  \delta \Gamma_{\alpha \beta}
\approx
  \sum_j g_{\alpha j} g_{j \beta}
  \frac {\rho_\chi}{m^2_V} \frac 1 {E_\nu}
  \gamma_0 \,.
\end{equation}
Consequently, the correction from vector dark matter to neutrino oscillation
takes the same form as the scalar case \geqn{eqS:HM} with
$\delta M^2_{\alpha \beta} = \mp
  2 \sum_j g_{\alpha j} g^*_{j \beta} \frac {\rho_\chi}{m^2_V}$,
which is similar as the scalar case with the Yukawa couplings $y$
replaced by the gauge couplings $g$.

\end{document}